\documentclass[A4,10pt]{article}

\usepackage{algorithm}
\usepackage{algpseudocode}  
\usepackage{esvect}
\usepackage{minted}
\usepackage{amsmath}
\usepackage{amssymb}
\usepackage{graphics}
\usepackage{hyperref}

\hypersetup{
	colorlinks=true,
	linkcolor=violet,
	filecolor=magenta,      
	urlcolor=cyan,
	citecolor=blue,        
	pdfpagemode=FullScreen,
	pdfauthor={Nicolas Olivier},
}
\RequirePackage[table]{xcolor}
\RequirePackage{tabularx,multirow,array}
\RequirePackage{graphicx,caption}

\begin{document}

\title{Modeling Coherent Nonlinear Microscopy of Axially Layered Anisotropic Materials Using FDTD}

\author{Mohammad Reza Farhadinia$^1$   and Nicolas Olivier$^{1,}$*}

\date{1 - Laboratoire d’Optique et Biosciences \\ École polytechnique, CNRS, Inserm, Institut Polytechnique de Paris, 91120 Palaiseau, France \\
\vspace{1cm} email: nicolas.olivier@polytechnique.edu}

\maketitle

\begin{abstract}
Providing quantitative interpretation of coherent nonlinear microscopy images, such as third-harmonic generation (THG), is generally hampered by the complex phase-matching conditions, especially in the presence of sample linear heterogeneity. We recently presented a numerical pipeline using the finite-difference time-domain (FDTD) method to take this heterogeneity into account. However, due to software restrictions, we only considered nonlinear materials with diagonal nonlinear susceptibilities.  We now expand the recently developed FDTD approach to model nonlinear microscopy for anisotropic materials that obey Kleinman Symmetry, organized in layers along the optical axis, and validate our simulations on well-described geometries.
\end{abstract}

\section*{Introduction}

Third-harmonic generation (THG) microscopy~\cite{barad1997nonlinear, muller19983d,yelin1999laser} is a coherent nonlinear microscopy modality increasingly used for the label-free characterization of cells and tissues ~\cite{oron2004depth, sun2004higher, debarre2006imaging,olivier2010cell, farrar2011vivo, weigelin2016third} often in combination with second-harmonic generation (SHG)~\cite{campagnola2003second} or with two or three-photon excited fluorescence~\cite{horton2013vivo}. 
Most biological materials have non-zero third-order nonlinear properties~\cite{debarre2007quantitative}, but the THG efficiency under high numerical aperture (NA) focusing conditions is principally determined by the presence of an axial $\pi$ phase shift of the excitation beam called the Gouy phase shift, and  when a beam is focused in a homogeneous, isotropic, normally dispersive medium with constant nonlinear properties, this phase shift prevents phase matching between fundamental and harmonic fields over the focal region, resulting in no detected THG ~\cite{ward1969optical,boyd2020nonlinear, barad1997nonlinear, cheng2002green}.
Therefore, THG images generally reveal interfaces or optical heterogeneities with sizes of a few hundred nanometers and highlight them over a dark background. In particular, lipid-water interfaces~\cite{debarre2006imaging, debarre2007quantitative, weigelin2016third} have been identified as efficient sources of contrast. Moreover, the visibility of specific structures depends not only on their optical properties but also on the interplay between the sample microstructure (size, shape, orientation) and the excitation field distribution (NA, Gaussian or higher order mode, aberrations)~\cite{cheng2002green, debarre2005structure, olivier2008third,olivier2012third,thayil2010influence}. In order to better understand THG images, a complete numerical framework has been developed, using the angular spectrum representation (ASR) to describe the tightly focused excitation beam,  nonlinear tensors in a specific arrangement that couple this excitation near focus to a nonlinear polarization source term, and propagation of this induced polarization to the detector plane using Green's functions~\cite{cheng2002green, sandkuijl2013numerical, olivier2012third}. This semi-analytical framework relies on the assumption that  there are no changes in the  linear optical properties of the medium considered. However, it has been identified that the presence of micron-scale refractive index heterogeneities has important consequences for CARS~\cite{van2016effects} and THG image contrast~\cite{morizet2021modeling,morizet2023third}. These effects occur in most biological samples and cannot be accounted for in ASR/Green models, outside of a few simple situations. We have recently shown that the finite-difference time-domain (FDTD) method could b e used in this situation. We established the validity of this approach for THG imaging by analyzing the polarized THG (PTHG) contrasts on an index-mismatched interface, spheres, cylinders~\cite{morizet2021modeling,morizet2023third}, and recently demonstrated the applicability of the approach to model the PTHG response of myelinated neurons \cite{morizet2025multiscale}. The main limitation of these studies was the use of diagonal nonlinear tensors, which meant only linear polarizations along $x$ or$y$ could be considered accurately, and which was due to software limitations. We now extend our model to anisotropic tensors in geometries consisting of layers along the optical axis and demonstrate {its} applicability to modeling THG in isotropic and anisotropic materials, as well as second-harmonic generation (SHG) in geometries consisting of materials layered along the optical axis. We also show that this approach can be extended to non-degenerate nonlinear processes such as sum-frequency generation (SFG) and third-order sum-frequency generation (TSFG)~\cite{ferrer2023label}, and could therefore in principle be used for any coherent microscopy method.

\section{Methods}

\subsection{Description of the FDTD model}

We performed FDTD simulations \cite{taflove2005computational, ccapouglu2012microscope, gallinet2015numerical} of THG by elaborating on the approach described in~\cite{morizet2021modeling,morizet2023third,morizet2025multiscale}. To limit the number of relevant parameters, we neglected linear dispersion by keeping linear indices constant. We worked with a commercial FDTD implementation (Ansys Lumerical FDTD, Synopsys Inc., Sunnyvale, CA). In this implementation, the electric and magnetic fields are calculated on every point of a 3D grid at successive times by solving discretized Maxwell equations for non-magnetic materials ($\vv{B} = \mu_0 \vv{H},\ \vv{D} = \varepsilon_0 \varepsilon_r \vv{E}$). The simulations are performed in the time domain, and spectral information can be retrieved using Fourier transforms. 

In the {Lumerical software} formalism, an arbitrary nonlinear polarization term can be introduced by calculating $\vv{D}$  explicitly as follows:

\begin{equation}{\vv{D}(\omega)=\varepsilon_{0} \vv{E}(\omega)} + \vv{P}(\omega)
\end{equation}
where $\vv{P}(\omega)$ is a function of $\vv{E}(\omega)$ defined in a material-dependent manner. For second and third harmonic generation microscopy, the response of a material can be described by $\chi^{(2)}$ and $\chi^{(3)}$ nonlinear tensors such that for $(i\in\{x,y,z\})$:
\begin{equation}P_i(t) = \varepsilon_0 \sum_{j}\chi^{(1)}_{ij} E_j(t) + \varepsilon _0 \sum_{j,k}\chi^{(2)}_{ijk} E_j(t)E_j(t)+ \varepsilon _0 \sum_{j,k,l}\chi^{(3)}_{ijkl} E_j(t)E_k(t)E_l(t) \,\,\, 
\end{equation}
A naive tensorial implementation would require computing $3^n$ combinations of electric fields for a n$^{th}$ order process, but assuming Kleinman symmetry, we can reduce significantly the number of required computations: for second-order nonlinear processes, we can write \cite{boyd2020nonlinear}:
\begin{equation} \begin{bmatrix}
P ^{(2)}_{x} \\
P^{(2)}_{y} \\
P^{(2)}_{z}
\end{bmatrix}
=2\begin{bmatrix}
d_{11} & d_{12} & d_{13} & d_{14} & d_{15} & d_{16}\\
d_{21} & d_{22} & d_{23} & d_{24} & d_{25} & d_{26}
\\ d_{31} & d_{32} & d_{33} & d_{34} & d_{35} & d_{36}
\end{bmatrix} \begin{bmatrix}
E^{2}_{x} \\
E^{2}_{y} \\
E^{2}_{z} \\
 2E_{y}E_{z}\\
 2E_{x}E_{z}\\
 2E_{x}E_{y} \\
\end{bmatrix} 
\label{eqshgtensor}
\end{equation}

And for a third-order process, we get:
\begin{equation}
\label{eqthgtensor}\begin{bmatrix}
P ^{(3)}_{x} \\
P ^{(3)}_{y} \\
P^{(3)}_{z}
\end{bmatrix}
=2\begin{bmatrix}
T_{11} & T_{12} & T_{13} & T_{14} & T_{15} & T_{16} & T_{17} & T_{18} & T_{19} & T_{110}\\
T_{21} & T_{22} & T_{23} & T_{24} & T_{25} & T_{26} & T_{27} & T_{28} & T_{29} & T_{210}
\\ T_{31} & T_{32} & T_{33} & T_{34} & T_{35} & T_{36} & T_{37} & T_{38} & T_{39} & T_{310}
\end{bmatrix} \begin{bmatrix}
E^{3}_{x} \\
E^{3}_{y} \\
E^{3}_{z} \\
 3E^{2}_{z}E_{x}\\
 3E^{2}_{z}E_{y}\\
 3E^{2}_{y}E_{z}\\
 3E^{2}_{y}E_{x}\\
 3E^{2}_{x}E_{y} \\
 3E^{2}_{x}E_{z} \\
6E_{x}E_{y}E_{z}
\end{bmatrix} 
\end{equation}
This means we need to compute 6 components for SHG and 10 for THG, which we can then combine into three nonlinear polarization terms, one per polarization (See Supplementary materials for a full description of the equations used in the different materials). We created four different types of nonlinear materials using the same principle:
\begin{enumerate}
    \item A second-order nonlinear material that includes terms from~\autoref{eqshgtensor}
    \item A third-order nonlinear material that includes terms from~\autoref{eqthgtensor}
    \item A second- and third-order nonlinear material that includes terms from both~\autoref{eqshgtensor} and~\autoref{eqthgtensor}
    \item An isotropic third-order nonlinear tensor that only requires one input argument, to avoid having to input the 30-element matrix every time.
\end{enumerate}

\subsection{Limitations of the anisotropic FDTD Model}

The nonlinear anisotropic material plugin we presented relies on the fact that the $x$, $y$, and $z$ components of the fields are stored in vectors that are ordered similarly (ie, the $n^{th}$ element of each vector corresponds to the same mesh point in the simulation). If this assumption holds, then we can simply multiply the electric field value vectors stored in the memory to calculate the nonlinear polarization in all mesh points. This is not always the case, as the software optimizes each polarization independently, and controlling its meshing behavior for complex geometries can be difficult, and will be the subject of future work. We therefore decided to focus on axially-layered structures, in which we found that the default meshing algorithm fulfilled our common ordering condition. Importantly, testing whether the mesh is appropriate before running the simulations is possible, and the code to test that is provided by Synopsys, Inc~\cite{Custommaterials}. 

Another fundamental limitation is that since the $E_x$, $E_y$ and $E_z$ components of the electric field are calculated in different positions in the Yee cell~\cite{taflove2005computational}, the nonlinear polarization represents a spatial average, which can lead to poor approximation of very small structures with strong near-field contributions, as commonly found in nanophotonics applications. However, in biological nonlinear microscopy applications we do not expect this to be an issue.  

Finally, the current GPU implementation of Ansys Lumerical FDTD introduces too much noise in the spectra, and is not suitable for nonlinear simulations (whether using the existing nonlinear materials, or the anisotropic materials presented here). Therefore, all simulations were run on the CPU.

\section{Results}

\subsection{THG of isotropic materials}

In order to test the validity of our approach, we first simulated a well-known condition. In the case of THG,  no signal can be obtained from a homogeneous isotropic material and the signal is limited to interfaces and inclusions between two different materials~\cite{cheng2002green}. One of the well-described consequences of the tensorial properties of THG is that the THG signal from a specific geometry (for example, an interface orthogonal to the optical axis between two materials with different values of $\chi^{(3)}$) vanishes when switching from linearly to circularly polarized light~\cite{oron2003depth}. Indeed, if we consider $\chi^{(3)}$ for an isotropic materials, we have~\cite{boyd2020nonlinear}:

\begin{equation}
\chi^{(3)}_{ijkl} = \chi_0 . (\delta_{ij}. \delta_{kl} + \delta_{ik}.\delta_{jl} +\delta_{il}.\delta_{jk})
\label{eq:ch3chi3ijkl} 
\end{equation}
which can be written using reduced tensor notations as:
\begin{equation}
\chi^{(3)}=\chi_0
\begin{bmatrix}
3 & 0 & 0 & 1 & 0 & 0 & 1 & 0 & 0 & 0 \\
0 & 3 & 0 & 0 & 1 & 0 & 0 & 1 & 0 & 0 \\
0 & 0 & 3 & 0 & 0 & 1 & 0 & 0 & 1 & 0
\end{bmatrix}
\end{equation}

The third-order nonlinear polarization can then be written as:

\begin{equation}
{\bf {P}}^{(3)} = 3\chi_0 \left[ \begin{array}{c}  E_x(E_x^2+E_y^2+E_z^2)  \\ 
 E_y(E_x^2+E_y^2+E_z^2)\\ E_z(E_x^2+E_y^2+E_z^2)  \end{array} \right]
\label{eq:p32} 
\end{equation}

Where the $E_z$ component of the field is quite low at this NA for a linearly polarized beam~\cite{cheng2002green,novotny2012principles} and can be almost neglected. This means that for a circularly polarized beam, we have 
\[E_x\approx iE_y \Rightarrow \,\,\,\, E_x^2+E_y^2\approx 0  \Rightarrow \,\,\,\, P_x^{(THG)}\approx P_y^{(THG)}\approx 0\]

\begin{figure}[htb]
\centering
\captionsetup{width=.8\linewidth}
\includegraphics[width=0.9\linewidth]{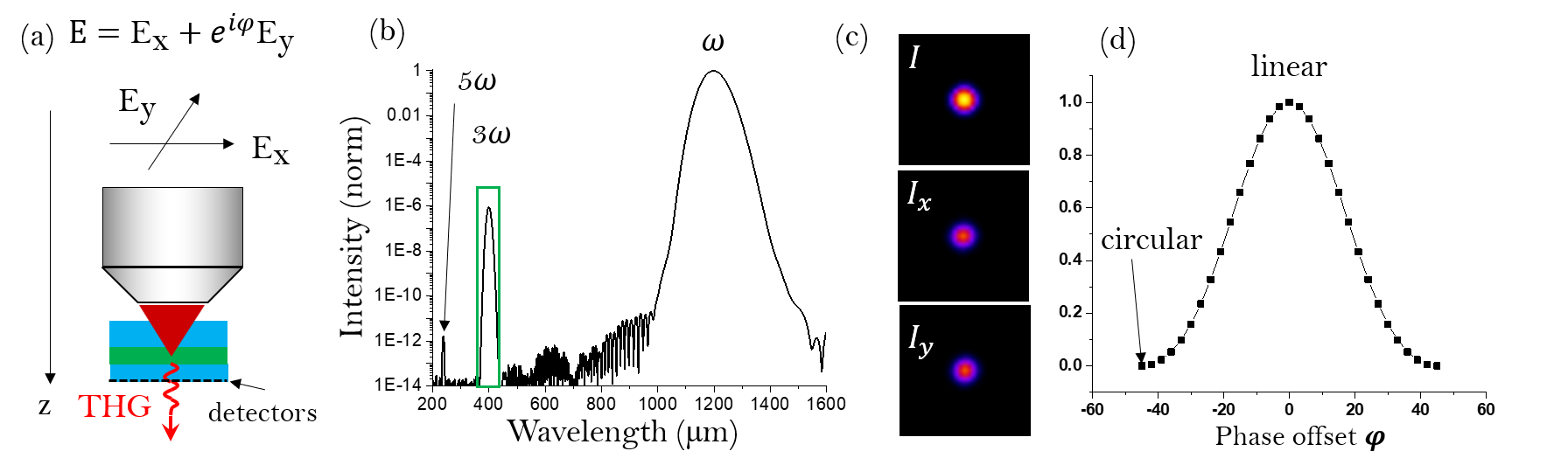} 
\caption{THG on a slab of isotropic $\chi^{(3)}$ material in water as a function of the polarization state. (a) geometry considered (b) Spectrum measured in the simulation with $\varphi=0$ (linear polarization) at the center of the detector array, showing the fundamental frequency and the THG signal. In green, the region over which the signal is integrated. (c) Intensity of the third harmonic field on the detector array, with from top to bottom total intensity, intensity along the x polarization, and along the y-polarization. (d) Simulated integrated THG as a function of the phase between the two polarizations.}
\label{fig:FigTHG}
\end{figure}

In order to simulate this condition, we considered a tightly focused (NA = 0.95) Gaussian beam at a central wavelength of $1.2$\,$\mu$m focused in the middle of a $2$\,$\mu$m slab of glass-like material with isotropic third-order nonlinear properties ($\chi_0=2$ [A.U.]) surrounded by a water-like material, also with isotropic third-order nonlinear properties but with lower nonlinear index ($\chi_0=1.68$ [A.U.]) (\autoref{fig:FigTHG}.a). We first consider an incoming linear polarization at $45$ degrees. We place a detector array far from the nonlinear material, and isolate the THG signal from the detected spectrum (\autoref{fig:FigTHG}.b). In this spectrum, we can also identify a weak contribution at $5\omega$, corresponding to cascaded THG, indicated by an arrow. Though we do detect this signal, which shows that the materials behave in the expected nonlinear manner, the meshing parameter we used in this simulation is probably too large to get an adequate quantitative estimate of what happens for 5HG.  Focusing on THG,  we can then reconstruct the emission profile of the THG field by plotting the intensity detected on the detector array, and we observe that the signal is emitted as predicted along the optical axis (\autoref{fig:FigTHG}.c), with an equal component polarized along $\mathbf{e_x}$ and $\mathbf{e_y}$ reflecting the isotropic response to the incoming polarization. We can then integrate this detected THG signal, and compare the signal obtained for different polarizations. We considered an elliptically polarized beam, where the $x$ and $y$ polarization are of equal amplitude, but there is a varying phase offset $\varphi$ between the two polarizations. When $\varphi=0$, we have a linearly polarized beam (along the direction $\mathbf{e_x}+\mathbf{e_y}$), while for $\varphi=90$ the beam is circularly polarized. The resulting simulated THG signal, a function of the angle $\varphi$ is given in \autoref{fig:FigTHG}.d and exhibit a clear maximum for linear polarization, with a vanishing amount of THG when the polarization is circular (both right-handed and left-handed circular polarization). 

\begin{figure}[hbt]
  \captionsetup{width=.8\linewidth}
\centering
\includegraphics[width=0.99\linewidth]{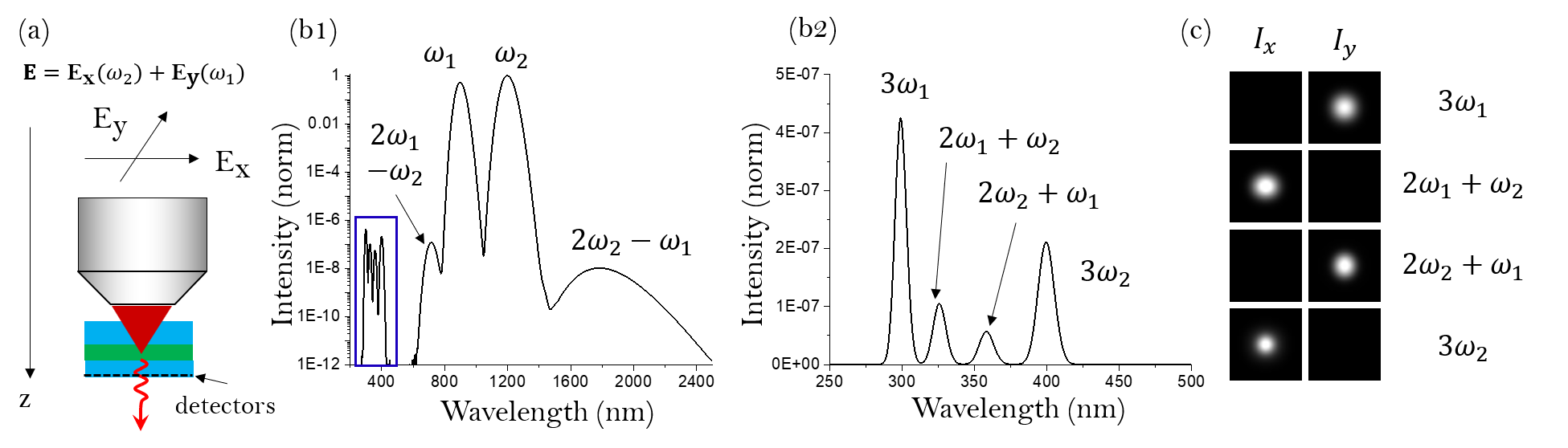} 
\caption{Non-degenerate third order nonlinear processes in a slab of  isotropic $\chi^{(3)}$ material  as a function of the polarization state.  (a) geometry considered (b1) Spectrum measured in the simulation at the center of the detector array shown on a log scale, showing the fundamental frequency, the two THG signals, the two 4-wave mixing signals at $2\omega_1-\omega_2$ and $2\omega_2-\omega_1$, and the 2 TSFG signals at $2\omega_1-\omega_2$ (b2) zoom on the region highlighted in blue in (b1) where we the four third-order sum-frequency signals using a linear scale, with 4 signals at $3\omega_1$, $2\omega_1+\omega_2$, $\omega_1+2\omega_2$, and $3\omega_2$.  (c) Intensity of the THG and TSFG fields on the detector array, along the $x$ polarization (left), and along the $y$ polarization (right) }
\label{fig:FigTSFG}
\end{figure}

 Using the same geometry, we then tested if we could model non-degenerate nonlinear processes such as FWM~\cite{mahou2011combined} or TSFG~\cite{ferrer2023label}, and therefore decided to consider two beams at two different wavelengths ($\lambda_1$=\,900\,nm, and $\lambda_2$=\,1.2\,$\mu$m) with orthogonal incoming polarizations (the beam at $\lambda_1$ along $\mathbf{e_y}$, and the beam at $\lambda_2$ along $\mathbf{e_x}$) (\autoref{fig:FigTSFG}.a). At the end of the simulation domain, we could  detect six different nonlinear signals (\autoref{fig:FigTSFG}.b1 and b2):
 \begin{itemize}
 \item the 2 THG signals, at $\lambda_1/3$=\,300\,nm, and $\lambda_2/3$=\,400\,nm
  \item the 2 four-wave mixing signals centered at $720$~nm ($2\omega_1-\omega_2$) and 1800~nm   ($2\omega_2-\omega_1$) 
  \item the 2 nondegenerate TSFG signals at $327$~nm (corresponding to $2\omega_1+\omega_2$) and $360$~nm ($2\omega_2+\omega_1$).
 \end{itemize}
Moreover, we could identify the polarization state of the nonlinear signals, and for two orthogonally polarized beams we observe linear THG signals parallel to the polarization of the fundamental beam, and for the TSFG signal which comes from non-diagonal tensor elements, we observe the expected dependence with $2\omega_2+\omega_1$ parallel to $3\omega_1$ and $2\omega_1+\omega_2$ parallel to $3\omega_2$.

Indeed, if we consider for example the $2\omega_2+\omega_1$ process for $E(\omega_1)$ polarized along $\mathbf{x}$ and $E(\omega_2)$ polarized along $\mathbf{y}$, and neglecting the $z$-component of the electric field for simplicity we have:

\begin{equation}
{\bf {P}}^{(2\omega_2+\omega_1)} = 3\chi_0 \left[ \begin{array}{c}  E_x({\omega_1})\left(E_x^2({\omega_2})+E_y^2({\omega_2})\right)  \\ 
E_y({\omega_1})\left(E_x^2({\omega_2})+E_y^2({\omega_2})\right)\\ 0  \end{array} \right]\approx 3\chi_0 \left[ \begin{array}{c}  E_x({\omega_1})E_y^2({\omega_2}) \\ 
0\\ 0  \end{array} \right]
\label{eq:tsfgtensor} 
\end{equation}

Since third-order processes in isotopic materials were adequately reproduced, we then decided to test second-order nonlinear processes.

\subsection{Collagen Fiber as SHG test of Anisotropic Material}

Second-harmonic generation requires anisotropy at the molecular level due to symmetry considerations~\cite{boyd2020nonlinear}, so modeling SHG requires anisotropic tensors. To validate our model, we tested it on a well-studied geometry, that of an interface between layers of uniaxial materials, such as found in the stroma of the cornea~\cite{meek2024structural}. In each layer, if we consider an array of collagen fibrils all oriented along the same direction, which we will choose as the x-axis, the second-order nonlinear susceptibility can be described as~\cite{gusachenko2012polarization,gusachenko2013numerical,teulon2015theoretical,galante2025intrinsic}:

\begin{equation} \label{eq_chi-optical-aligned}
\begin{split}
\chi^{(2)}_{xxx} &= 1.36\chi^{(2)}_0 \\
\chi^{(2)}_{xyy} &= \chi^{(2)}_{yyx} =\chi^{(2)}_{yxy} = \chi^{(2)}_0 \\
\chi^{(2)}_{xzz} &= \chi^{(2)}_{zxz} =\chi^{(2)}_{zzx} = \chi^{(2)}_0  \\
\end{split}
\end{equation}

So, the nonlinear tensor can be expressed using the reduced tensor as:

\begin{equation}
\chi^{(2)}=\chi^{(2)}_0 
\begin{bmatrix}
1.36 & 1 & 1 & 0 & 0 & 0 \\
0 & 0 & 0 & 0 & 0 & 1 \\
0 & 0 & 0 & 0 & 1 & 0 
\end{bmatrix}
\label{eq:tensorcollagen}
\end{equation}
for a material-oriented along the $x$ direction.

\begin{figure}[ht]
  \captionsetup{width=.8\linewidth}
\centering
\includegraphics[width=0.99\linewidth]{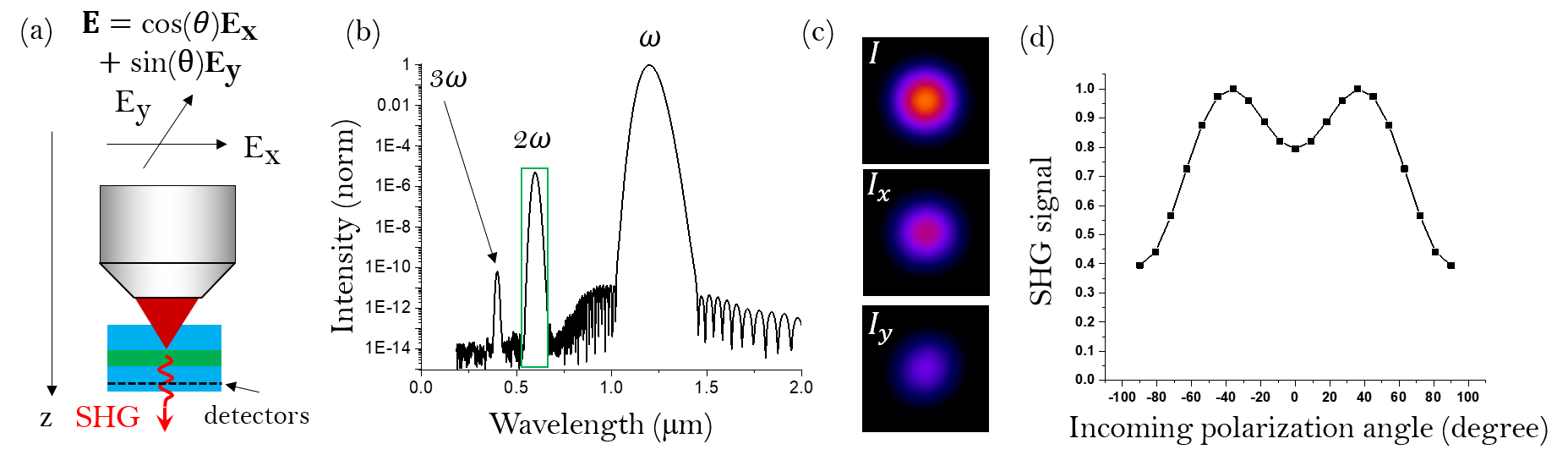} 
\caption{SHG in an anisotropic material mimicking a 2$\mu$m layer of the corneal stroma oriented along $x$ as a function of the incident linear polarization angle (a) geometry considered, with the nonlinear signal detected in transmission (b) Spectrum measured in the simulation at the center of the detector array shown in a log scale, where we can identify the fundamental frequency at 1.2$\mu$m ($\omega$), the SHG signal at 600~nm (2$\omega$) and a cascaded SHG peak at 400~nm (3$\omega$). In green, the spectral region over which the SHG signal is integrated. (c) The intensity of the SHG field on the detector array, with from top to bottom: the total intensity, the intensity along the $x$ polarization, and the intensity along the $y$ polarization (d) Integrated SHG intensity as a function of the linear polarization angle from the $x$ axis (so 0 is $x$ polarized, 90 is $y$ polarized)}
\label{fig:FigSHG}
\end{figure}

We considered again a tightly focused (NA = 0.95) Gaussian beam at a central wavelength of $\lambda_0= 1.2\,\mu$m focused in the middle of a $2$~$\mu$m slab, but this time the slab had second-order nonlinear properties described by~\autoref{eq:tensorcollagen}, and the surrounding material only has linear properties (\autoref{fig:FigSHG}.a). We first considered a fundamental beam polarized linearly along $x$, and in the detector closest to the optical axis we could identify in the detected spectrum a clear SHG peak at $\lambda_0/2 =$ 600\,nm, as well as some cascaded SHG signal giving rise to a weak signal at $\lambda_0/3=$ 400\,nm (\autoref{fig:FigSHG}.b). Plotting the integrated SHG signal as a function of detector position and polarization allows us to determine the SHG emission profile, which is in the case also along the optical axis, but with a larger divergence than in the THG signal simulated previously. (\autoref{fig:FigSHG}.c). The presence of a $y$ component for the nonlinear signal for an incoming $x$ polarized field is the signature of the presence of non-diagonal terms in the nonlinear tensor. Finally, we repeated this simulation by changing the angle of the incoming linear polarization, and plotted the integrated SHG intensity as a function of the angle in~\autoref{fig:FigSHG}.d.  This predicted polarization dependence is in line with experimental measurements~\cite{aptel2010multimodal,raoux2021quantitative} and other numerical approaches~\cite{gusachenko2012polarization,gusachenko2013numerical}.

\begin{figure}[hb]
  \captionsetup{width=.8\linewidth}
\centering
\includegraphics[width=0.99\linewidth]{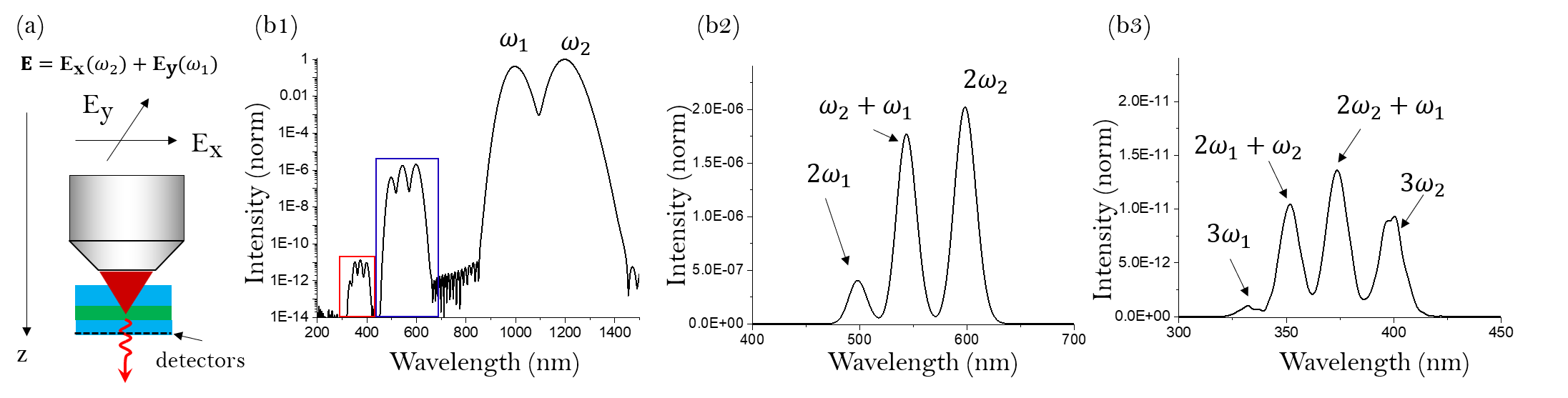} 
\caption{SHG and SFG in an anisotropic material mimicking the layer of the corneal stroma (a) geometry considered, with two incoming Gaussian beams at $\lambda_1=1000$\,nm polarized along $x$ and $\lambda_1=1200$\,nm polarized along $y$.  (b1) Spectrum measured in the simulation at the center of the detector array on a log scale, showing the fundamental frequencies, and the nonlinear signals. In blue, a region corresponding to second-order processes displayed on a linear scale in (b2), where we can identify 3 peaks corresponding to the 2 SHG signals and the SFG signal, and in red to cascaded processes displayed on a linear scale in (b3) with 4 peaks corresponding to $3\omega_1$,$2\omega_1+\omega_2$,$\omega_1+2\omega_2$, and $3\omega_2$}
\label{fig:Fig4SFG}
\end{figure}

We then ran a new simulation where we added another beam centered at a wavelength $\lambda_0=$1000~nm,  with a linear polarization orthogonal to the 1200~nm beam (\autoref{fig:Fig4SFG}.a). If we look at the detected spectrum we can see the presence a nondegenerate signal corresponding to sum-frequency generation (SFG)  at 550\,nm ($\omega_1+\omega_2$)(\autoref{fig:Fig4SFG}.b). Moreover, we can even see very weak cascaded SFG signals, which provide some weak signals at 4 additional frequencies (\autoref{fig:Fig4SFG}.c) corresponding to the 4 different way to add up 2 different frequencies. Taken together, \autoref{fig:FigSHG} and \autoref{fig:Fig4SFG} show that the approach presented is suitable to model both degenerate and nondegenerate second-order nonlinear processes. We can therefore model both third-order processes and second-order processes, and we will see in the next section  how to model  materials that have both second and third-order nonlinear properties.

\subsection{Collagen Fibers as an Anisotropic source of both SHG and THG}
As a material that has both $\chi^{(2)}$ and $\chi^{(3)}$ nonlinear properties, we once again consider collagen fibers arrays as in the previous situation. Indeed, arrays of collagen fibers, for example those found in the stroma, are a known source of SHG, but we know from experiments that they can also generate some THG signal, and with a detected polarization dependence that is different from a cascaded second-order process~\cite{aptel2010multimodal}. Since we consider a uniaxial crystal, which we will assume is oriented along the $x$ axis, the induced nonlinear polarization  can be modeled as~\cite{olivier2010harmonic,boyd2020nonlinear}:
\begin{equation}
{\bf {P}}^{(3\omega)} =  \left[ \begin{array}{c}  E_x(\chi_{\parallel}.E_x^2+\chi_{cr}.E_y^2+\chi_{cr}.E_z^2)  \\ 
 E_y(\chi_{cr}.E_x^2+\chi_{\bot}.E_y^2+\chi_{\bot}.E_z^2)\\ E_z(\chi_{cr}.E_x^2+\chi_{\bot}.E_y^2+\chi_{\bot}.E_z^2)  \end{array} \right]
\label{eq:equation_cornea} 
\end{equation}
which means using the reduced tensors, we have:

\begin{equation}
\chi^{(3)}_{ijkl}
=
\begin{bmatrix}
\chi_{\parallel} & 0 & 0 & \chi_{cr}/3 & 0 & 0 & \chi_{cr}/3 & 0 & 0 & 0
\\
0 & \chi_{\bot} & 0 & 0 & \chi_{\bot}/3  & 0 & 0 & \chi_{cr}/3  & 0 & 0
\\ 
0 & 0 &\chi_{\bot} & 0 & 0 & \chi_{\bot}/3  & 0 & 0 & \chi_{cr}/3  & 0
\end{bmatrix}
\end{equation} 

\begin{figure}[htb]
  \captionsetup{width=.8\linewidth}
\centering
\includegraphics[width=0.99\linewidth]{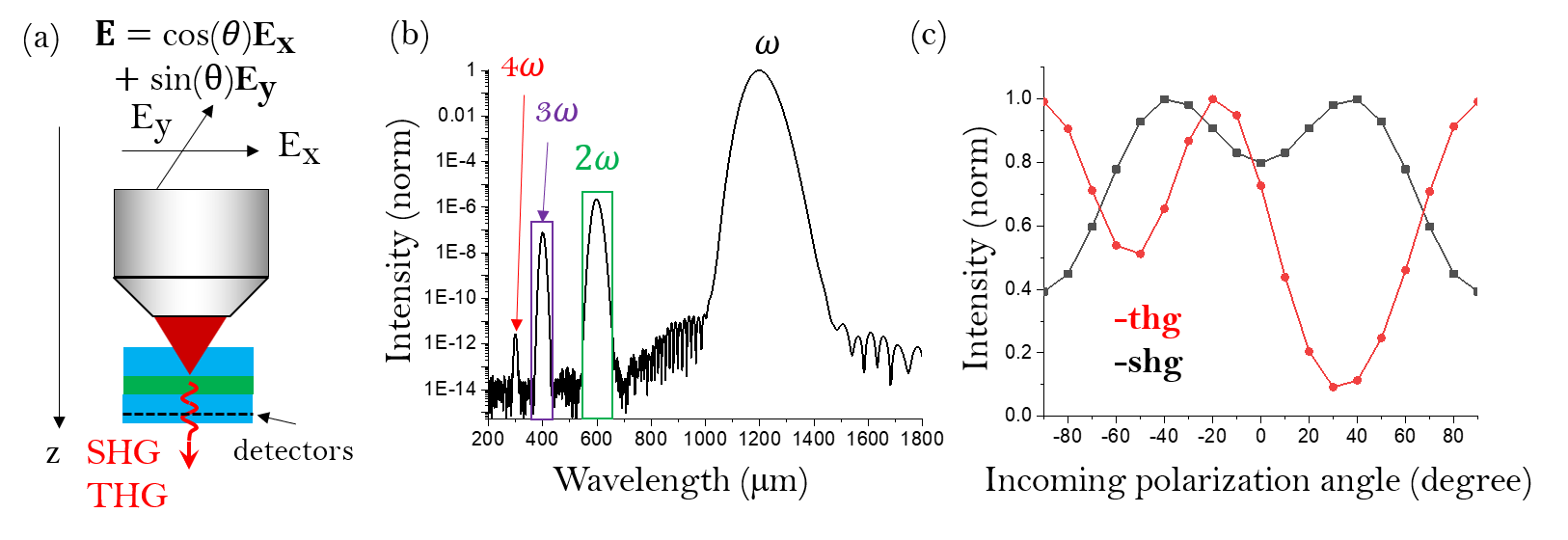} 
\caption{SHG and THG in an anisotropic material mimicking the layer of the corneal stroma as a function of the incident linear polarization angle (a) geometry considered (b) Spectrum measured in the simulation at the center of the detector array, showing the fundamental frequency, the SHG signal at 600~nm (and a weak cascaded SHG peak at 300~nm) and thg THG signal at 400~nm. In green, the region over which the SHG signal is integrated, and in purple, the region where THG is integrated. (c)  simulated integrated SHG and THG signal as a function of the linear polarization direction}
\label{fig:Fig5SHG+THG}
\end{figure}

To test the simulation, we decided to use $\chi_{\bot} = \chi_{cr}=\chi^{(3)}_0$, and $\chi_{\parallel}=1.5\chi_{\bot}$.  We decided to use the same geometry as in \autoref{fig:FigTHG}, with a value of zero for the $\chi_0^{(2)}$ of water, for the collagen layer the same second order tensor as in \autoref{fig:FigSHG} but we decreased the value of $\chi_0^{(2)}$. We again decided to simulate the dependence of the nonlinear signals as a function of the angle of the incoming polarization (\autoref{fig:Fig5SHG+THG}.a)

We first used an incoming linear polarization at $45$ degrees as we did for THG, and in this situation, we can see on the spectrum both the SHG signal at 400~nm and the SHG signal at 600~nm, as well as a weak cascaded 4HG signal (\autoref{fig:Fig5SHG+THG}.b).  When looking at the polarization dependence of the signal, we can see (\autoref{fig:Fig5SHG+THG}.c) that while the SHG signal at 600\,nm has an identical dependence as that of \autoref{fig:FigSHG}, which is expected since the presence of a third-order nonlinear process should not affect this, the signal at 400\,nm which consists of a sum of cascaded second order and direct third-order processes presents a highly non-trivial polarization dependence. The ratio between the different contributions (and therefore the overall THG response) depends on the intrinsic ratio between $\chi^{(2)}$ and $\chi^{(3)}$, which we have not determined experimentally. 

\section{Conclusion and Perspectives}
We present some new FDTD-based coherent nonlinear microscopy simulations that rely on a set of plugins to describe materials with non-diagonal second-order and third-order nonlinear tensors that obey Kleinman symmetry. These new materials plugins enable us to run FDTD simulations that quantitatively reproduce other simulation approaches for THG and TSFG in isotropic materials, as well as SHG and SFG from a collagen-rich structure similar to the corneal stroma. While we focus here on validating the numerical model, and therefore use materials that all share the same linear properties to get a common reference with the other models, we can use FDTD to study the influence of the linear indices of the different materials present in the focal volume. Moreover, by treating multiple nonlinear processes in parallel, this type of simulation could allow us to study the interplay between cascaded SHG and direct THG, for example, in materials such as calcite. 

The implementation we present depends on insuring the mesh is ordered properly, which can be done using a check function, and we show it is robust when looking at axially layered geometries for both degenerate and non-degenerate second-and third order nonlinear processes.  Further work on optimizing the meshing process under suitable constraints should make this approach generally applicable for all geometries in coherent nonlinear microscopy.

\section*{Acknowledgments}
We thank Federico Duque Gomez from Synopsys, Inc. for his technical support on the material plugin feature in Ansys Lumerical FDTD, as well as Romin Durand for his work on a preliminary implementation of the plugin for nonlinear simulations. We thank the members of the Advanced Microscopies group at LOB for their discussions on nonlinear microscopy, in particular Emmanuel Beaurepaire for THG/TSFG and Marie-Claire Schanne-Klein for SHG/SFG.

\section*{Disclosures}
The authors declare no conflicts of interest.

\section*{Data availability} {The scripts and the code used to compile the material dlls  used to run the simulations underlying the results presented in this paper are available upon request for now, and will be uploaded to Zenodo upon publication.
 A detailed description of how to compile and use the new material plugins to run new simulations is available at \url{https://farhadinia.github.io/Nonlinear-Anisotropic-Material-Plugin/}}

\bibliographystyle{unsrt}
\bibliography{biblio}

\end{document}